\documentclass[a4paper,fleqn,usenatbib]{mnras}

\usepackage[T1]{fontenc}
\usepackage{ae,aecompl}

\usepackage{graphicx}	
\usepackage{amsmath}	
\usepackage{amssymb}	

\usepackage{color}

\title[Four dimensional map]{Chaotic orbits obeying one isolating integral
in a four dimensional map}

\author[J. C. Muzzio]
{J. C. Muzzio$^{1,2}$,\thanks{E-mail: jcmuzzio@fcaglp.unlp.edu.ar}\\
$^{1}$Facultad de Ciencias Astron\'omicas y Geof\'isicas, Universidad Nacional
de La Plata, Paseo del Bosque, 1900 La Plata, Argentina\\
$^{2}$Instituto de Astrof\'isica de La Plata (CCT CONICET La Plata--UNLP),
Paseo del Bosque, 1900 La Plata, Argentina}

\date{Accepted XXX. Received YYY; in original form ZZZ}

\pubyear{2017}

\begin{document}
\label{firstpage}
\pagerange{\pageref{firstpage}--\pageref{lastpage}}
\maketitle

\begin{abstract}
We have recently presented strong evidence that chaotic orbits
that obey one isolating integral besides energy exist in a toy Hamiltonian model
with three degrees of freedom and are bounded by regular orbits that isolate
them from the Arnold web. The interval covered by those numerical experiments
was equivalent to about one million Hubble times in a galactic context. Here we
use a four dimensional map to confirm our previous results and to extend that
interval fifty times. We show that, at least within that
interval, features found in lower dimension Hamiltonian systems and maps
are also present in our study, e.g., within the phase space occupied by
a chaotic orbit that obeys one integral there are subspaces where that orbit does
not enter and are, instead, occupied by regular orbits that, if tori, bound other
chaotic orbits obeying one integral and, if cantori, produce stickiness. We argue
that the validity of our results might exceed the time intervals
covered by the numerical experiments.

\end{abstract}

\begin{keywords}
physical data and processes: chaos -- methods: numerical -- galaxies: kinematics and
dynamics -- celestial mechanics
\end{keywords}



\section{Introduction}

   Autonomous Hamiltonian systems with three degrees of freedom can,
in principle, support three types of orbits: regular that obey two
isolating integrals besides energy, partially chaotic that obey only
one integral besides energy, and fully chaotic that obey just the energy
integral. Nevertheless, it is still unclear whether partially chaotic
orbits actually exist. While examples of those orbits have been presented
by \citet{Cont1978} and by \citet{PV1984}, their existence has been
denied by \citet{F1970}, \citet{F1971} and \citet{LL1992}. Full details on
this situation are given by \citet{M2017}, who also found
partially chaotic orbits that are bounded by regular orbits in a toy
model; the validity of his results is however limited to the time span
covered by his numerical investigation that, although very long (about
one million Hubble times if placed in a galactic context), is not infinite.
As indicated by \citet{M2017}, if partially chaotic orbits actually exist,
they can pose obstacles to chaotic diffusion: three dimensional (3-D hereafter)
regular orbits cannot bound fully chaotic orbits that are 5-D, but they can
bound 4-D partially chaotic orbits and these, in turn, can either bound or
place obstacles to the fully chaotic orbits. It is thus important from a
theoretical point of view to establish whether partially chaotic orbits
exist or not.

   Partially chaotic orbits are also of interest for dynamical astronomy.
In our own work on the dynamics of triaxial stellar systems that goes
from \citet{M2003} to \citet{CM2016} (see the latter for references to
other works) we have shown that partially chaotic orbits represent about
10 per cent of the orbits in those systems and have a distribution that
differs from that of the fully chaotic orbits. Thus, it is necessary
to distinguish partially from fully chaotic orbits in studies on the
dynamics of elliptical galaxies. The presence of partially chaotic orbits
in those systems had been noticed earlier by \citet{GS1981} and \citet{MV1996},
but no particular importance was given to them at that time. Besides, the
phenomenon of 'stable chaos' investigated by \citet{MN1992} and \citet{Mil1997}
might be an example of partially chaotic orbits in the Solar System.

It will be worthwhile, before going on, to clarify the
meaning of some terms to avoid confusion. The terms partially and fully chaotic
orbits were proposed by us \citep[][]{MCW2005} to design what \citet{PV1984} had
called, respectively, weakly and strongly chaotic orbits. The reason was that,
as indicated by \citet{C2002}, the terms weak and strong chaos are used by
other authors \citep[see, e.g.,][]{VCE1998} to refer to the value of the
Lyapunov exponent (LE hereafter) and that is the meaning that we prefer for
those terms. Therefore, what \citet{PV1984} called weakly chaotic orbit is just
the same of what we call partially chaotic orbit. What most authors nowadays
call weakly chaotic orbit is an orbit with a low value of its largest LE (the only
one that is usually computed). But a partially chaotic orbit is one that has its
largest LE different from zero, no matter whether it is large or small, and its
second largest LE equal to zero (see Subsection 2.2 for details).

The case of sticky orbits is different. They are orbits that stay
for a long time in a certain region of phase space and, then, diffuse into another
region. One example are orbits that stay long around islands of stability, another
are orbits close to the unstable asymptotic curves of unstable periodic orbits.
They were first noted by \citet{C1971}, and later on by \citet{SR1982}, 
\citet{K1983} and \citet{Meiss1983}; Karney seems to have been the first one
to use the term sticky in this context. For reviews and more recent results
see \citet{CH2010a} and \citet{CH2010b}. Sticky orbits play an important role
in barred spiral galaxies because they support the shape of the bar, as well as the
rings and spiral arms, for very long times before escaping through Arnold diffusion
\citep[see, e.g.,][]{CH2013, HK2009}. For a recent and detailed study of the speed
of that diffusion see \citet{EH2013}.
Stickiness is a phenomenon different from partial and full chaos and, in principle,
it can affect both partially and fully chaotic orbits, although thus far there were
no known examples because most authors do not make the distinction between partially
and fully chaotic orbits in their studies. We will present here, however, an example
of stickiness in a partially chaotic orbit (at least within the interval covered
by our numerical iterations).

   As explained by \citet{M2017} it would be very difficult to extend further
his investigation on a toy Hamiltonian model, but an excellent opportunity
is offered instead by 4-D maps. These maps are a prototype of the Poincar\'e map
that one obtains with a cut through the phase space of an autonomous Hamiltonian
system with three degrees of freedom and they do not demand the numerical
integration of differential equations. In fact, the results of \citet{F1971}
were obtained with a map, a pionnering study that he continued in \citet{F1972},
\citet{FS1973a} and \citet{FS1973b}. But, excellent as it is, his work was
limited by the computational means available at that time and, in particular,
it did not reach the degree of resolution needed to find the very fine regions
where \citet{M2017} showed that may lurk the partially chaotic
orbits. Therefore we decided, resorting to the means nowadays available, to
investigate the possible existence of partially chaotic orbits in the same map
he had studied and that is the subject of the present paper. Besides
providing strong evidence that partially
chaotic orbits exist in the map studied by Froeschl\'e, the present investigation
extends that of \citet{M2017} in two ways. First, it yields similar results using
a map instead of a Hamiltonian model and, second, it extends the validity of
those results (limited to the interval covered by the numerical integrations or
iterations) to a 50 times longer interval.
We used the same techniques we had employed in our investigation of the toy model,
adapting them to the case of maps. We also made good use of 3-D plots using color
as the fourth dimension, a technique developed by \citet{PZ1994} to investigate
Hamiltonian systems with three degrees of freedom \citep[see
also][for a more recent application to a case that includes stickiness]{Kats2013}.
A similar technique was applied by \citet{Rich2014} to the case of maps.

   The organization of this paper is very similar to that of our previous one.
The following section describes the map and the techniques we used to study
its orbits. The results of a search for possible partially chaotic orbits are
presented in Section 3, where we also isolate one of them using the
integral it obeys. In section 4 we present the regular orbits that bound that
partially chaotic orbit, and show 3-D Poincar\'e maps with those three orbits
plus a few other interesting ones and, finally, we explain our conclusions in
Section 5.

\section{Map and numerical methods}

\subsection{The map}
\label{map} 

We chose the 4-D map:
\begin{equation}
    x_{i+1} = x_{i} + a_{1}~sin(x_{i}+y_{i}) + b~sin(x_{i}+y_{i}+z_{i}+t_{i})
	\label{mapx}
\end{equation}
\begin{equation}
    y_{i+1} = x_{i} + y_{i}
	\label{mapy}
\end{equation}
\begin{equation}
    z_{i+1} = z_{i} + a_{2}~sin(z_{i}+t_{i}) + b~sin(x_{i}+y_{i}+z_{i}+t_{i})
	\label{mapz}
\end{equation}
\begin{equation}
    t_{i+1} = z_{i} + t_{i}
	\label{mapt}
\end{equation}
where the values of $x, y, z$ and $t$ lie always between $-\pi$ and $+\pi$ and
the determinant of its Jacobian matrix is equal to 1.
As indicated in the Introduction this model was extensively investigated by
\citet{F1971}, \citet{F1972}, \citet{FS1973a} and \citet{FS1973b}.
For the present investigation we adopted $a_{1} = a_{2} = -0.25$ and $b = 0.02$
which yield plots similar to those of the double resonance studied by \citet{M2017}.

\subsection{Numerical methods}
\label{methods} 

We explored the phase space of our model using orbits with initial coordinates
$z=t=0$ and different $x$ and $y$ values.
To follow each orbit and at the same time compute the four LEs 
we adapted for our map the {\sc liamag} routine, kindly
provided by D. Pfenniger \citep[see][]{UP1988} and originally written for orbits
in a Hamiltonian system. This was simply done replacing the call to the
Runge-Kutta-Fehlberg subroutine that integrated the orbits and the variational
equations by the map equations \ref{mapx} through \ref{mapt} and by the Jacobian
matrix of the map. We prepared two versions of the routine,
one using double precision as the original version, and a second one using quadruple
precision. The latter allowed us to follow the orbits with much longer
iterations but, of course, it run much slower.

The LEs $\lambda_{1} > \lambda_{2} > \lambda_{3} > \lambda_{4}$  have the
property that $\lambda_{i}=-\lambda_{5-i}$, because the determinant of the Jacobian
matrix is equal to 1. Besides, each isolating integral makes zero one
$\lambda_{i}=-\lambda_{5-i}$ pair so that, considering only the two largest
LEs, we have that both are zero for regular orbits, only one is zero
for partially chaotic orbits, and none is zero for fully chaotic orbits.
Nevertheless, since the number of iterations computed for the map is
necessarily finite, numerical LEs can tend towards zero as the number of iterations
in the map increases, but they remain always larger than a limiting value that can be
estimated to be of the order of $ln N/N$, where $N$ is the number of iterations.
This is a coarse estimate only, and the limiting value should be determined in every
case \citep[see, e.g.,][for details]{ZM2012}.

\begin{figure}
	\includegraphics[width=\columnwidth]{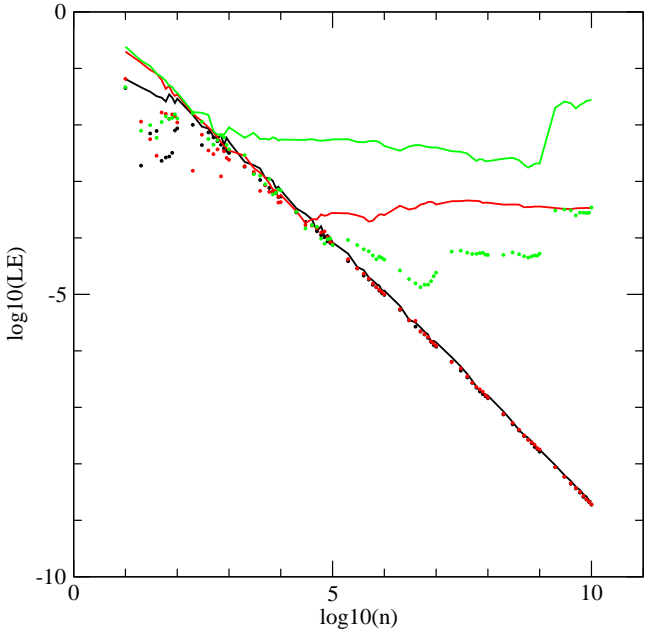}
    \caption{Evolution of the LEs with the number of iterations
for regular, partially chaotic (red in the electronic version) and fully chaotic
(green in the electronic version) orbits. For each orbit $\lambda_{1}$ is shown
with full lines and $\lambda_{2}$ with dots.}
    \label{LEs}
\end{figure}

Figure~\ref{LEs} gives the evolution of the LEs of three
different orbits as the number of iterations increases. For each orbit full
lines were used for $\lambda_{1}$ and dots for $\lambda_{2}$. The regular
orbit r1 (see Subsection 4.1) is shown in black, the partially chaotic orbit
pch (see Subsection 3.1) in red in the electronic version and the fully chaotic
orbit with initial conditions $x = 0.87500, y = 2.015625, z = 0, t = 0$ in green
in the electronic version. We notice that, for the regular orbit, both LEs decrease
almost linearly with the number of iterations in this double logarithmic plot,
and that the same happens with the $\lambda_{2}$ value of the partially chaotic
orbit. The $\lambda_{1}$ value of the partially chaotic orbit, as well as
both LEs of the fully chaotic orbit, instead, clearly do not go to
zero as the number of iterations increase.

The longest computations done with the double precision routine reached $10^{8}$
iterations and a comparison with the results of the quadruple precision routine
showed that the errors in $x, y, z$ and $t$ were at most of the order of $10^{-5}$.
With the quadruple precision routine we reached up to $10^{10}$ iterations and a
comparison with results obtained reversing the iteration and using the inverse map
suggested that the errors in $x, y, z$ and $t$ were at most of the order of
$2 \times 10^{-7}$. Taking our map as a Poincar\'e map, each iteration in the former
corresponds to the interval elapsed between two cuts to obtain the latter, i.e., we
can take an iteration as the characteristic time of the orbits in the Hamiltonian
that corresponds to the Poincar\'e map. Thus our $10^{10}$ iterations can be taken
as covering an interval of $10^{10}$ characteristic times and, since 1 Hubble
time corresponds to about 200 characteristic times for the elliptical galaxies we
investigated previously \citep[see, e.g.,][]{ZM2012}, that is equivalente to about
$5 \times 10^{7}$ Hubble times in a galactic context.

\begin{figure}
	\includegraphics[width=\columnwidth]{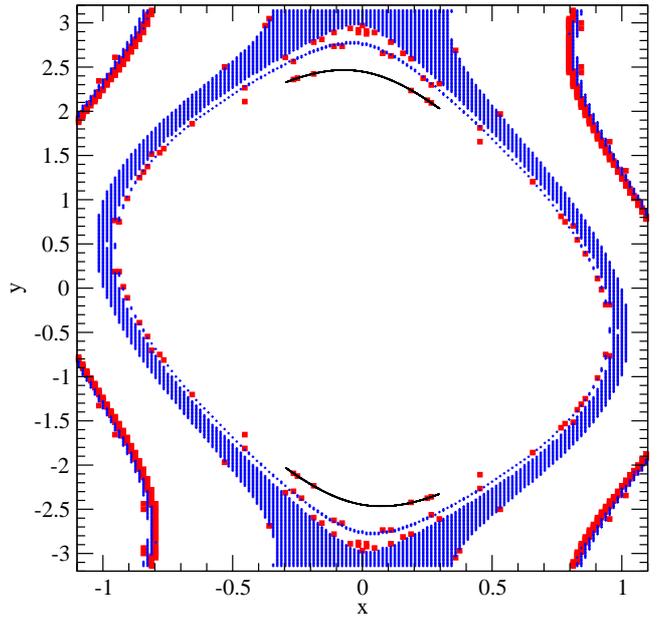}
    \caption{Initial conditions on the $(x, y)$ plane of
orbits classified as regular, partially and fully chaotic from
the values of their LEs.
The blank areas correspond to regular orbits, partially chaotic orbits are
shown as filled squares (red in the electronic version) and fully chaotic
orbits as plus signs (blue in the electronic version). The small dots that
trace the two curves above and below were obtained taking two slices
$\vert z \vert \leq 5. \times 10^{-5}$ and $\vert t \vert \leq 5. \times 10^{-5}$
from the partially chaotic orbit pch (see Subsection 3.2 for details).}
    \label{wholexy}
\end{figure}

Orbits that obey no integral fill in a 4-D $(x, y,z, t)$ space and those that obey
one integral, $I_{1}(x,y,z,t) = C_{1} = constant$, occupy a 3-D space because, in
principle, we might put, e.g., $t$ as a function of $x, y, z$ and $C_{1}$. Besides,
orbits that also obey a second integral, $I_{2}(x,y,z,t) = C_{2} = constant$,
occupy a 2-D space because, in principle, we might also put, e.g., $z$ as a function
of $x, y, C_{1}$ and $C_{2}$. Thus, in order to recognize regular from
chaotic orbits we need to take a cut, say, $z = z_{o}$ to get curves for the regular
orbits and surfaces for the chaotic ones. In practice, the cut has to be replaced by a
slice $z \simeq z_{o}$ to get a reasonable number of points, but the width of the
slice can be kept thin enough to avoid affecting the recognition of the orbits. It is
important to remember that, as indicated by \citet{M2017}, these surfaces and curves
do not lie on a plane but are warped because they are embbeded in a 3-D space.
A second slice, say $t \simeq t_{o}$, is needed to distinguish partially from fully chaotic
orbits: the former will appear as curves and the latter as surfaces (and regular orbits
as points).

Therefore, we prepared a quadruple precision program that gave the successive iterations
of the map \ref{mapx} through \ref{mapt} and took slices $z \simeq z_{o}$ and $t \simeq t_{o}$.
The results presented here were obtained with $10^{10}$ iterations and a
comparison with results obtained reversing the iteration and using the inverse map
suggested that the errors in $x, y, z$ and $t$ were at most of the order of $2.5 \times 10^{-7}$,
i.e., essentially the same as those of the {\sc liamag} routine for the same number of
iterations, as could be expected. As explained before, our $10^{10}$ iterations correspond
to about $5 \times 10^{7}$ Hubble times in a galactic context.

\begin{figure}
	\includegraphics[width=\columnwidth]{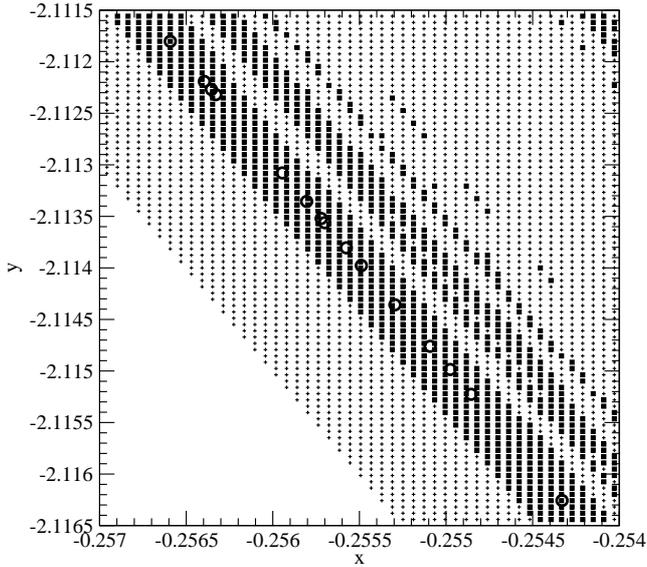}
    \caption{High resolution plot of a small section of Figure 1. Regular orbits
are shown as crosses and partially chaotic orbits as filled squares. The open
circles result from taking two slices $\vert z \vert \leq 5. \times 10^{-5}$ and
$\vert t \vert \leq 5. \times 10^{-5}$ from the same partially chaotic orbit of
Figure~\ref{wholexy}. The white area was not investigated with
this high resolution, but lower resolution plots showed only regular orbits there.}
    \label{lanexy}
\end{figure}

\begin{figure}
	\includegraphics[width=\columnwidth]{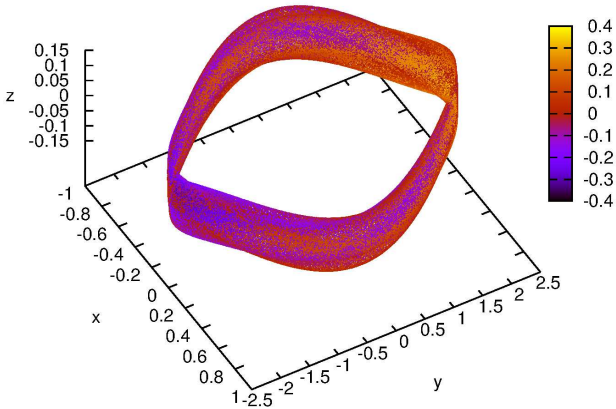}
    \caption{Partially chaotic orbit pch, from the chain shown in Fig. 2.
It is shown in the 3-D space $(x, y, z)$ with the fourth dimension $t$
given by the colour scale in the electronic version.}
    \label{torus}
\end{figure}

In order to get a clear view of orbital structures at several stages of our
investigation we found very useful the technique of \citet{PZ1994} who
used 3-D plots plus colour to represent the fourth dimension. We have adopted
their method using gnuplot (Copyright (C) 1986 - 1993, 1998, 2004, 2007 Thomas
Williams, Colin Kelley) to make the plots.

\begin{figure}
	\includegraphics[width=\columnwidth]{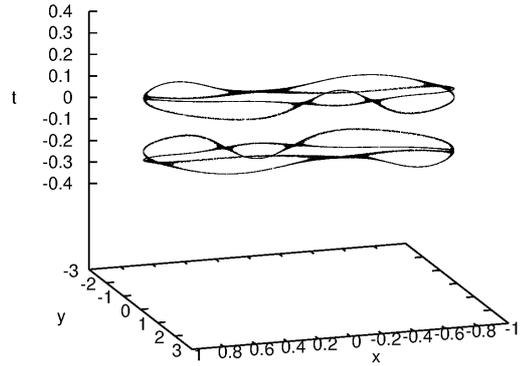}
    \caption{A slice $\vert z \vert \leq 10^{-6}$ of the partially
chaotic orbit pch from Figure 3 in the 3-D space $(x, y, t)$}
    \label{cut}
\end{figure}

To represent the warped surfaces and curves that result from our cuts we resorted
to the method of \citet{M2017} that used Fourier series to fit a chosen regular
orbit and to refer to it the results from nearby orbits. Here we adopted the
$x, y$ and $t$ variables resulting from the $\vert z \vert \leq 10^{-6}$ slice 
and, for an orbit selected as reference, we normalized each one of those variables 
subtracting the corresponding value for the center of the orbit and dividing the
result by the dispersion of the variable in question. Then we transformed that
normalized coordinate system into a cilindrical one and, using the azimuth angle
$\phi$ as argument, we obtained the best fitting Fourier series for the radius $R$
and the vertical distance $Z$ (notice that this new
$Z$ variable is actually the normalized value of the original $t$, not $z$,
variable). For nearby orbits, we obtained the differences between the
values of their variables (normalized using the same center and dispersions of the
orbit taken as reference) and those given by the corresponding Fourier series,
so that we can travel along the orbits, following $\phi$, and
the differences between their $R$ and $Z$ values and the corresponding
ones of the orbit taken as reference can be plotted with considerable detail. Here
we have adopted a cylindrical system of coordinates, rather than the spherical one
we used in our previous work, because the former is better to show the warped
surfaces we found in the present investigation.

\begin{figure}
	\includegraphics[width=\columnwidth]{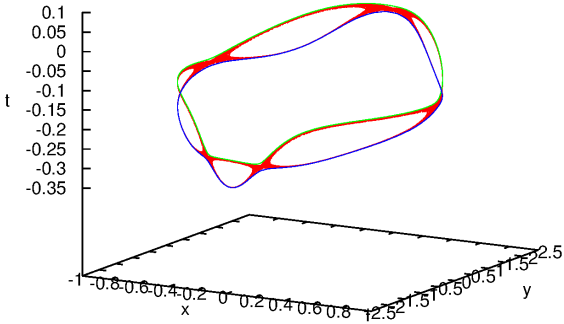}
    \caption{A slice $\vert z \vert \leq 10^{-6}$ of the partially
chaotic orbit pch from Figure 3 in the 3-D space $(x, y, t)$ (red in the
electronic version) together with the same slices from the two nearby regular
orbits r1 (green in the electronic version) and r2 (blue in the electronic version).
The slice has actually two parts as shown in Figure 4 but, for clarity, only the
part with lower $t$ values is shown here.}
    \label{cut3}
\end{figure}

We experimented with different numbers of terms and
found that the mean square error decreased as we increased that number up to
about 651 terms (that is, up to terms $sin(325\phi)$ and $cos(325\phi)$) and
reached a plateau where increasing the number of terms did not significantly
decreased the mean square error any further, so that we adopted that number
of terms for our computations. For slices with $\vert z \vert \leq 10^{-6}$
the resulting mean square errors of the $x$, $y$ and $t$ variables turned
out to be of the order of $0.15 \times 10^{-7}$,
$0.74 \times 10^{-7}$ and
$6.4 \times 10^{-7}$, respectively. For slices with double and half widths the
errors were proportional to the width of the slice, as could be expected. To
estimate the numerical errors of iteration we obtained the
Fourier series using only the first 20 per cent points and computed the
mean square errors of the last 20 per cent points with respect to those series.
The dispersions turned out to be essentially the same, so that
the errors of the numerical iteration should be much smaller than the
dispersion caused by the slice widths.

\section{Partially chaotic orbits}

The first step of our investigation was to search for possible partially
chaotic orbits in our map. We performed that search using LEs but,
as indicated by \citet{LL1992}, we can never be sure that the partially chaotic
orbits found in that way will not appear as fully chaotic with LEs computed
with a larger number of iterations. Thus, it should be recalled
that the orbits that we will refer to as partially chaotic here can be
regarded as such only over the span covered by our iterations.

\subsection{The search}
\label{energyplane}
We began our search preparing a sample of initial conditions
with $z=t=0$ and a grid of $x$ and $y$ values with
$-\pi < x < \pi$ and $-\pi < y < \pi$ and $2^{-6} = 0.015625$
spacing. The advantage of taking these initial
conditions is that the plots obtained with them will be useful as comparison
when, later on, we will take slices with $z \simeq 0$ and $t \simeq 0$. 
Using those initial conditions we computed the  orbits over $10^{7}$ iterations
and obtained the LEs which, in turn, we used to classify the orbits as regular,
partially or fully chaotic.
Figure~\ref{wholexy} is a $(x, y)$ plot where the blank areas correspond to
initial conditions that yielded regular orbits, while those that yielded
partially and fully chaotic orbits are shown, respectively, as filled squares
(red in the electronic version) and plus signs (blue in the electronic version).
The chains of small dots resulted from taking two slices $\vert z \vert \leq 5. \times 10^{-5}$
and $\vert t \vert \leq 5. \times 10^{-5}$ from a partially chaotic orbit, and will be
explained in Subsection 3.2. The Figure has several features in
common with Figure 1 of \citet{M2017}: we notice a central region
dominated by regular orbits, surrounded by another one dominated by fully chaotic
ones, with most of the partially chaotic orbits lying on the border between
those regions.

What most interests us here is that, as in our previous work, several partially
chaotic orbits can be found also well inside the regular domain. Therefore, we
made a higher resolution plot of the region $-0.265 < x < -0.250, -2.125 < y < -2.100$,
where a couple of partially chaotic orbits can be seen in Figure~\ref{wholexy}, that
showed a continuous chain of partially chaotic orbits as we had found before. We
did plots of increasingly higher resolution that showed the same and, besides, resolved
the chain in several parallel chains of partially chaotic orbits separated by similar
chains of regular orbits. The plots were obtained with $10^{8}$ iterations, but the
regular and partially chaotic nature of several of the orbits was confirmed running the
{\sc liamag} routine up to $10^{10}$ iterations. Figure~\ref{lanexy} shows a small part of
one of our plots obtained with a grid spacing of $2^{-14} \simeq 0.000061035$. Regular
orbits are shown as crosses and partially chaotic orbits as filled squares. We also show,
as open circles, the results from taking two slices $\vert z \vert \leq 5. \times 10^{-5}$
and $\vert t \vert \leq 5. \times 10^{-5}$ from a partially chaotic orbit, that will be
explained in Subsection 3.2. 

Figure~\ref{torus} shows, in the 3-D space $(x, y, z)$ and using colour to
represent the fourth dimension $t$, the orbit whose initial conditions are
$x=-0.256591796875, y=-2.11180224609375, z=0, t=0$ and that we will dub pch
hereafter. This is one of the partially chaotic orbits that lie on the chain shown in
Figure~\ref{lanexy} and its partially chaotic nature was confirmed by the LEs obtained
running the {\sc liamag} for $10^{10}$ iterations. It has the form of a torus and some
mixing of the colours might be present, a characteristic of chaotic orbits in this
sort of plot as indicated by \citet{PZ1994}, but if it exists it is far from clear. In
fact, except perhaps for the colour distribution, similar plots for other orbits
from the same region, either regular or partially chaotic, look very much the same.

Figure~\ref{cut} shows a slice $\vert z \vert \leq 10^{-6}$ from the orbit of
Figure~\ref{torus} in the 3-D space $(x, y, t)$ and we see that it has two parts, very
similar to each other, one with mainly positive $t$ values and another with mainly
negative $t$ values. For the time being, we will concentrate on the part that corresponds
to the lower values of $t$ which is shown again in Figure~\ref{cut3} \footnote{Notice
that  Figures~\ref{cut} and~\ref{cut3} show the 3-D space from two different points of
view and that is why, in the latter, the orbits seem to reach positive $t$ values,
it is just an effect of perspective.}. The Figure
shows the aforementioned slice from the orbit pch in the 3-D space $(x, y, t)$ (red in
the electronic version) together with similar slices from regular orbits r1 (green in
the electronic version) and r2 (blue in the electronic version), that will be explained
in Subsection 4.1. As anticipated, the points lie on a warped surface (actually, it has
a very small width because there is a finite range of $z$ values) and not on a plane.
The regular orbits are curves that bound the surface occupied by the chaotic orbit,
which is a double orbit (with each part similar to each one of the regular orbits)
linked by a bifurcation that is the most likely source of its chaos. All this is
very similar to what we found in \citet{M2017}.

\subsection{The integral of motion of partially chaotic orbits}
\label{integral}

Since the initial conditions we used to obtain Figures~\ref{wholexy} and~\ref{lanexy}
had all $z = t = 0$, these figures are similar to Poincar\'e maps resulting from
two cuts $z = 0$ and $t = 0$. But, for a true Poincar\'e map, we should have selected
orbits that have the same value of the integral that obey the partially chaotic orbits
(and one of the two integrals that obey the regular orbits), so as to get 1-D curves
for the partially chaotic orbits and points for the regular ones. Therefore, the lane of
partially chaotic orbits in Figure~\ref{lanexy} can be seen as the surface that results
from placing one beside another the curves corresponding to partially chaotic orbits
with different values of that integral, again a situation very similar to the one
found in \citet{M2017}, and our problem is to segregate the different curves that
make up the lane. But here we have the enormous advantage that we can obtain many
more points per orbit than we had in our previous work, so that
it is perfectly possible to make a long iteration of a partially chaotic
orbit and to take slices $z \simeq 0$ and $t \simeq 0$ thin enough to get the curve
it traces on the $(x, y)$ plane.

We selected the partially chaotic orbit pch, iterated it $10^{10}$ times and obtained the slices
$\vert z \vert \leq 5. \times 10^{-5}$ and $\vert t \vert \leq 5. \times 10^{-5}$.
The result is shown in Figures~\ref{wholexy} and~\ref{lanexy} where we notice that
we obtained a curve that is very thin indeed confirming that it corresponds to a
partially chaotic orbit as we had already found with the LEs. One minor difference with
the result obtained by \citet{M2017} is that here the curves corresponding to constant
values of the integral run parallel to the lane, while in the toy Hamiltonian model
they crossed it (cf. Figure 8 of our previous paper).

\section{Bounding regular orbits}
\subsection{Finding the boundaries}
\label{boundaries}
 
The task of finding regular orbits that obey the same integral as partially chaotic orbit
pch and bound it is also greatly simplified thanks to the large number of points that we
have at our disposal here. We can simply extrapolate the lines obtained from the two
slices in $z$ and in $t$, rather than having to resort to surfaces as done by
\citet{M2017}. Extrapolations are always risky and non linear extrapolations are
the riskiest, so that we decided to take only a small section from the right tip of
the lower curve of Figure~\ref{wholexy} given by the slices
$\vert z \vert \leq 5. \times 10^{-5}$ and $\vert t \vert \leq 5. \times 10^{-5}$
and to perform a linear extrapolation. Of course, as the tori of the regular orbits
fit one inside the other, one can choose from the extrapolation many points that
correspond to different tori that share the same value of the integral with the partially
chaotic orbit pch. We chose one with $x = 0.30250000, y = -2.31987691, z = 0.0, t = 0.0$
that, without demanding much extrapolation, provides the intial conditions of the
regular orbit we dubbed r1 that yields clear plots. The dispersion of the points around
the extrapolating straight line was $2.5 \times 10^{-6}$, further proof that we are dealing
with a very thin line, and the estimated error of the $y$ value of the extrapolated point
was $5.0 \times 10^{-6}$. To get an estimate of how this result is affected by the fact
that the line is not perfectly straight, we fitted one line to the first half of points
and another to the second half and the $y$ difference between the two extrapolations at
$x = 0.30250000$ turned out to be $6.6 \times 10^{-5}$, a precision more than enough for
our purposes.

The slices $\vert z \vert \leq 5. \times 10^{-5}$ and $\vert t \vert \leq 5. \times 10^{-5}$
from orbit r1 produce points on the extrapolation of the left tip of the same curve, so that
it cannot be used to find the other bounding orbit as we had expected. The reason is that this
second bounding orbit has no points near $z = t = 0$, but this problem was easily solved
taking from the partially chaotic orbit pch two new slices $\vert z \vert \leq 5. \times 10^{-5}$
and $\vert t+0.265 \vert \leq 5. \times 10^{-5}$. Another linear fit and extrapolation to the
tip of the resulting curve let us find the initial conditions for the regular orbit that
we dubbed r2 at $x = 0.17220000, y = -2.39949868, z = 0.0, t = -0.265$.

Figure~\ref{cut3} shows that, indeed, partially chaotic orbit pch is bounded by the regular
orbits r1 and r2. But a clearer view can be obtained with the technique developed by
\citet{M2017} to get 3-D Poincer\'e maps, that offers 2-D plots rather than the 3-D one
shown in the Figure, and that will be the subject of the next Subsection.

The surface that results from taking the slice  $\vert z \vert \leq 5. \times 10^{-5}$
from the orbit pch has holes in it, as can be seen in Figure~\ref{cut3}, so that we also
searched for orbits inside those holes to include them in our 3-D Poincar\'e maps. Taking
from orbit pch two new slices $\vert z \vert \leq 5. \times 10^{-5}$
and $\vert t+0.230 \vert \leq 5. \times 10^{-5}$ and performing another linear extrapolation
we found the partially chaotic orbit phol and the regular orbit rhol whose initial conditions
are, respectively, $x = 0.26465000, y = -2.34147742, z = 0.0, t = -0.230$ and
$x = 0.26342500, y = -2.34259592, z = 0.0, t = -0.230$.

The regular or partially chaotic nature of all the orbits found in the present subsection
was confirmed with runs of $10^{10}$ iterations with the {\sc liamag} routine.

\subsection{3-D Poincar\'e maps}
\label{maps}

We chose the part with lower $t$ values that results from taking the slice
$\vert z \vert \leq 10^{-6}$ from orbit r1 as our reference orbit and we computed the mean
values ($<x>, <y>$ and $<t>$) and the dispersions ($\sigma_{x}, \sigma_{y}$ and $\sigma_{t}$),
of its $x$, $y$ and $t$ values. Taking those mean values as the
center of the orbit, we computed the normalized values $(x-<x>)/\sigma_{x},
(y-<y>)/\sigma_{y}$ and $(t-<t>)/\sigma_{t}$ and used these normalized
values to define a new cilindrical system of coordinates, with azimuth angle
$\phi$, radius $R$, and vertical distance $Z$ (recall that this $Z$ is just
the normalized value of $t$ and has nothing to do with $z$). Then, taking $\phi$
as argument, we adjusted each normalized coordinate with a Fourier series
and we used them to iteratively improve the center of the orbit. Finally,
we obtained new Fourier series to represent $R$ and $Z$
as functions of $\phi$. The differences between the
true values and those given by the series, i.e. the residuals, were
used to represent orbit r1 in our 3-D Poincar\'e maps, that is,
straight lines with some dispersion through $R = 0$ and $Z = 0$,
respectively.

\begin{figure}
\centering
\includegraphics[width=7.6cm]{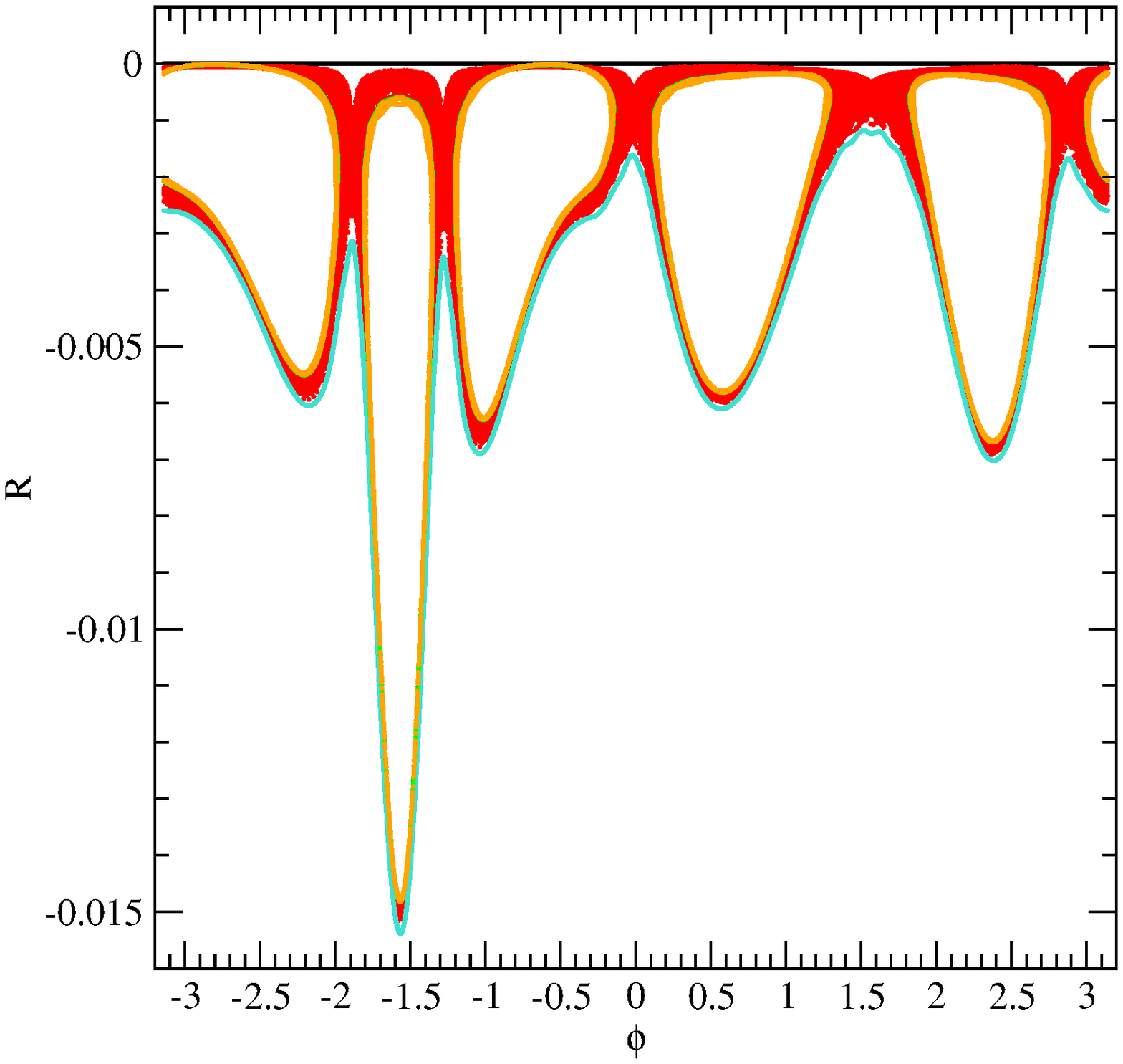}
\vskip 5.mm
\includegraphics[width=7.6cm]{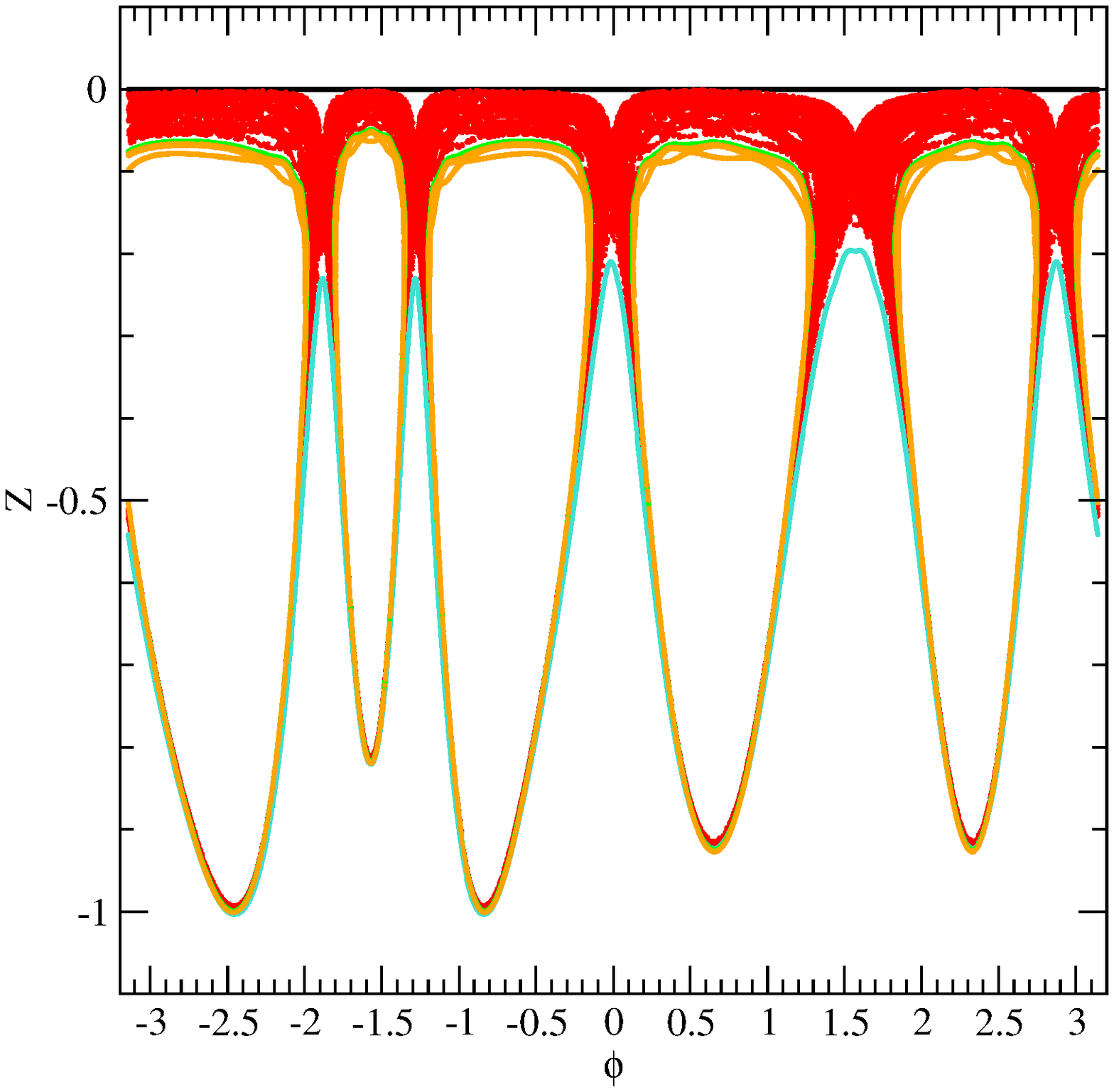}
    \caption{3-D Poincar\'e maps $(\phi, R)$ (above) and $(\phi, Z)$
(below) of orbits r1, r2 (turquoise in the electronic version, pch (red in
the electronic version), rhol (green in the electronic version) and
phol (orange in the electronic version). They correspond to the
$\vert z \vert \leq 10^{-6})$ slice and to the part of the orbits with lower $t$
values. The ordinates give the differences between
the values of $R$ and $Z$, respectively, of each orbit and those given by the
Fourier series fitted to the correponding values of orbit r1. See text for explanation.}
    \label{pmaplow}
\end{figure}

For other orbits we normalized their $x, y$ and
$t$ values using the same center and dispersions adopted for the r1 orbit
and obtained the corresponding $\phi, R$ and $Z$ values.
Finally, using their $\phi$ values as argument of the Fourier series
obtained for r1, we obtained the differences between their $R$
and $Z$ values and those given by the series. In other words,
our 3-D Poincar\'e maps are just the differences between each orbit
and r1, so that we can clearly represent those small differences as
we follow the orbit through all the different azimuth angles.

\begin{figure}
\centering
\includegraphics[width=7.6cm]{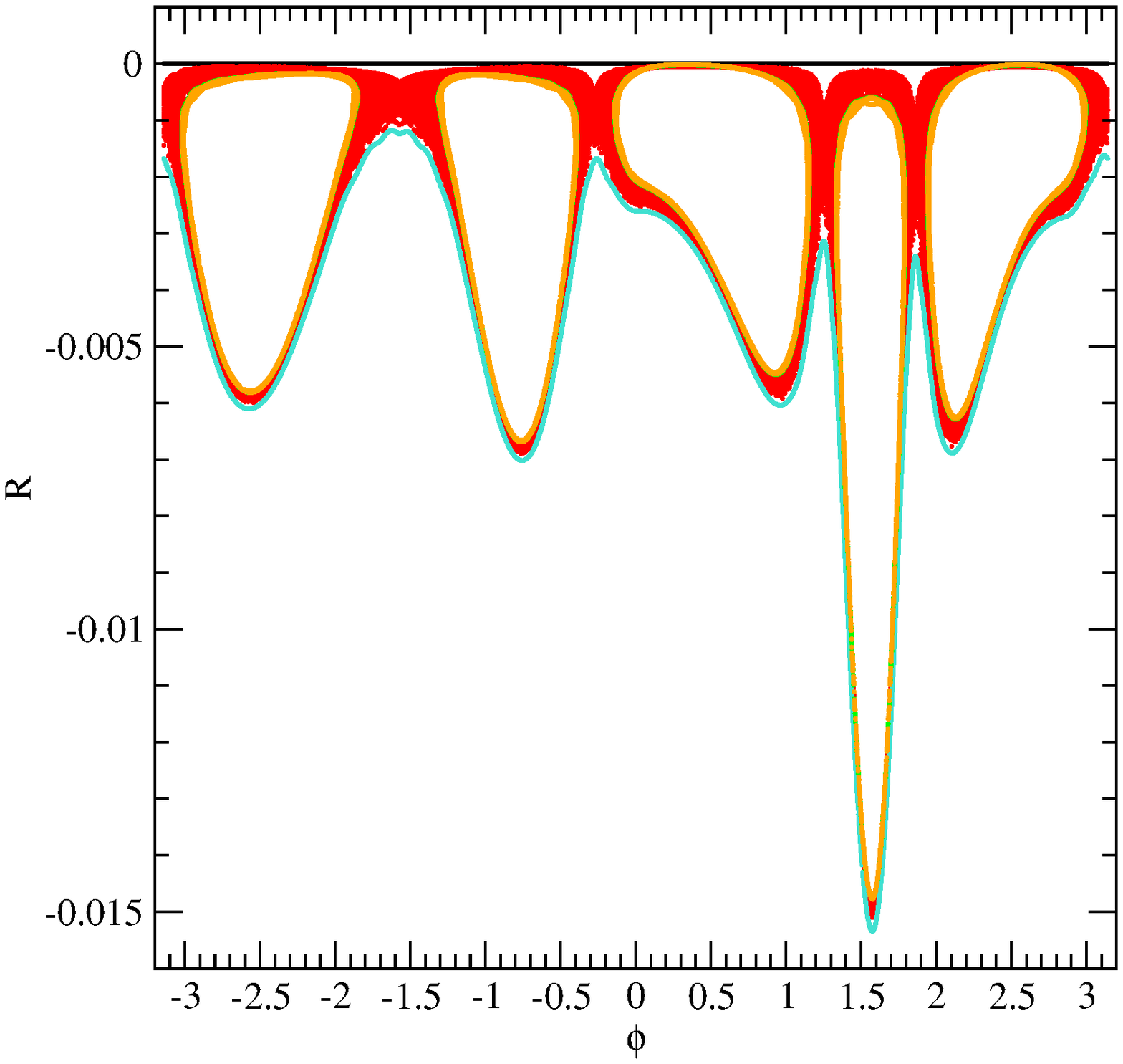}
\vskip 5.mm
\includegraphics[width=7.6cm]{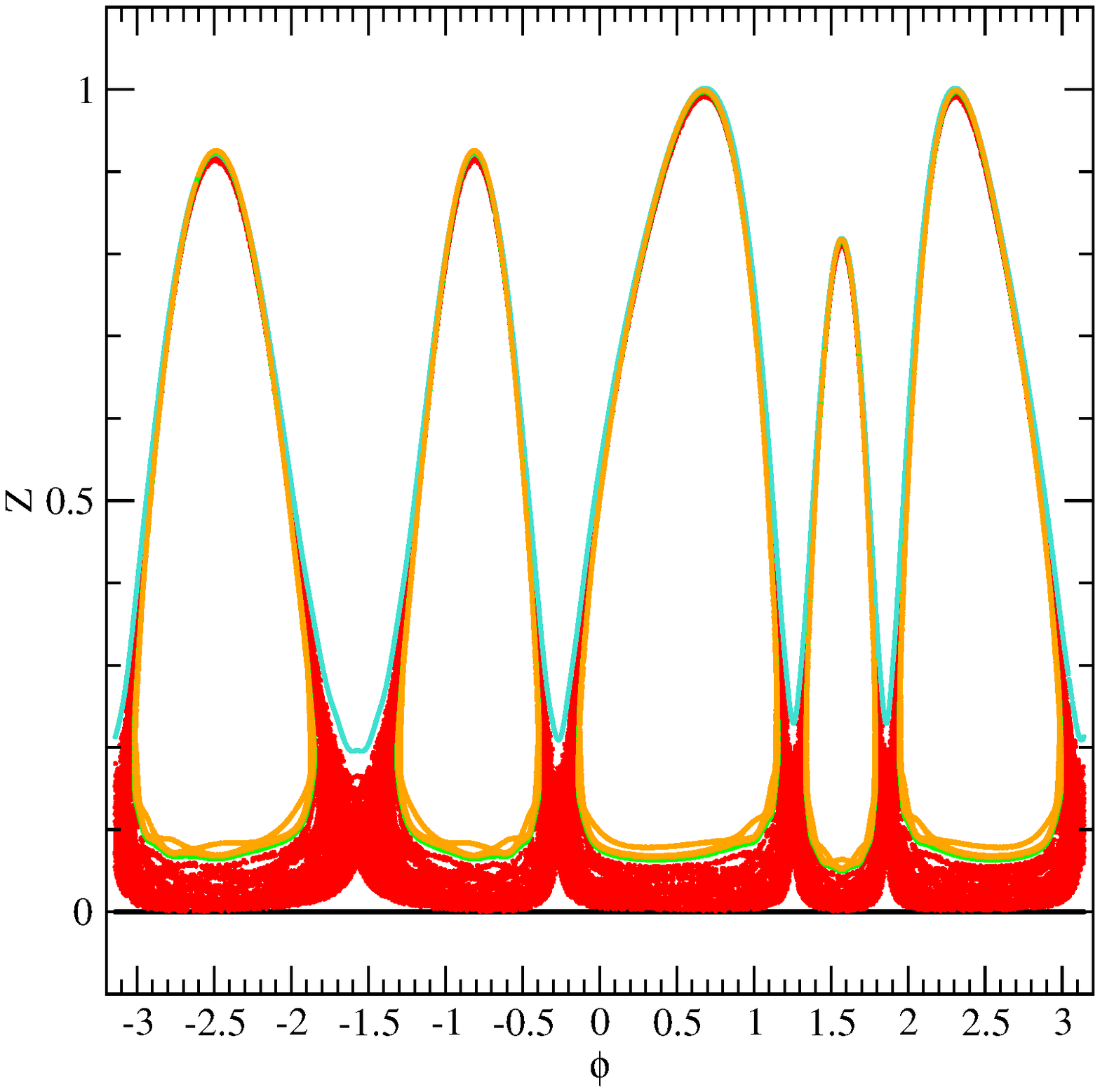}
    \caption{Same as Figure 7, but for the part of the orbits with higher $t$ values.}
    \label{pmaphigh}
\end{figure}

Figure~\ref{pmaplow} presents the result and, since the slices are warped, there seems to
be some crossing among the different orbits, but this is only apparent. The $\phi$ versus
$R$ plot (above) clearly shows that the regular orbits bound the partially chaotic orbits
everywhere except for $R$ values very close to zero but, at the same time, the $\phi$ versus
$Z$ plot (below) clearly shows that the regular orbits bound the partially chaotic orbit
for the corresponding $Z$ values. That is, the crossing of orbits takes place at different
values of $\phi$ in each plot, so that there is no actual crossing in 3-D space.

Of particular interest is the fact that inside partially chaotic orbit pch, and separated
from it by regular orbit rhol, lies partially chaotic orbit phol. Therefore, we not only have
the partially chaotic orbit pch well isolated from the rest of the phase space by regular orbits
r1 and r2, but it even has inside it the partially chaotic orbit phol well protected by the
cocoon provided by regular orbit rhol. As could be expected, the largest
LE of orbit phol is lower ($3.5 \times 10^{-5}$) than that of orbit pch ($3.4 \times 10^{-4}$);
as a comparison, for $10^{10}$ iterations, the LEs of regular orbits, and also the lowest LE
of those partially chaotic orbits, are about $2. \times 10^{-9}$. 

Figure~\ref{pmaphigh} is like Figure~\ref{pmaplow}, but for the part of the orbits with
higher $t$ values. Although we had not used that part of orbit pch to find the bounding
regular orbits and those inside its holes, this new Figure tells the same story as the
previous one: partially chaotic orbit pch is bounded by regular orbits r1 and r2 and
contains inside its hole partially chaotic orbit phol separated from pch by regular orbit
rhol.
 
\subsection{Stickiness}
\label{stick}

\begin{figure}
\centering
\includegraphics[width=7.6cm]{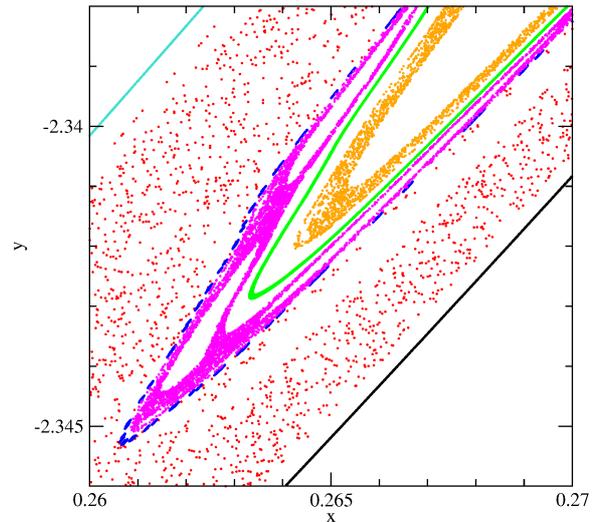}
    \caption{x vs. y plot of the slice $\vert z \vert \leq 10^{-5}$ showing part of the
hole in the orbit pch (red in the electronic version), together with orbits r1, r2
(turquoise in the electronic version), rhol (green in the electronic version), phol
(orange in the electronic versioin), cant (blue in the electronic version), and pchb
(magenta in the electronic version).}
    \label{xycant}
\end{figure}

When we were looking for the bounding regular orbits, we perfomed a few experiments
fitting planes to small sections of the $\vert z \vert \leq 10^{-5}$ slice
of orbit pch. It was soon clear that the extrapolations done in that way could
not reach distances as long as those obtained fitting a line to the slices
$\vert z \vert \leq 5. \times 10^{-5}$ and $\vert t \vert \leq 5. \times 10^{-5}$ and
which was the method adopted here. Nevertheless, in the process we found the partially
chaotic orbits pchb ($x = 0.26269580, y = -2.34256250, z = 0.0, t =-0.23280000$) and
pchc ($x = 0.26313362, y = -2.34209380, z = 0.0, t =-0.23280000$), and the regular
orbit cant ($x = 0.262461560, y = -2.34271880, z = 0.0, t =-0.23280000$).
Their regular or partially chaotic nature was confirmed running the {\sc liamag}
routine for $10^{10}$ iterations.
Figure~\ref{xycant} shows, on the $(x, y)$ plane the slice $\vert z \vert \leq 10^{-5}$
of orbits pch, pchb and cant (the same slice for orbits r1, r2, rhol and phol were
also added for comparison). We notice that partially chaotic orbits pch, pchb and
phol occupy areas while regular orbits r1, r2 and rhol are curves, the first and the
second ones bounding orbit pch and rhol separating pchb from phol. But regular orbit
cant is not a continuous but a broken line, i.e., it is a cantori that cannot separate
orbits pch and pchb. In fact, we found that the area covered by orbit pchc (not shown)
superposes with the areas covered by both pch and pchb, so that these two orbits are
in fact a single one. In other words, we have here an example of the phenomenon of
stickiness \citep[see, e.g.][]{C2002} where
orbit cant (and others like it) presents a barrier to the motion between the regions
of phase space covered by orbits pch and pchb, but porous enough to be occasionally
traversed. Since the region covered by pch is much larger than that covered by pchb
(notice the big difference between the density of points on each area) it is much
less likely that the barrier posed by the cantori could be traverse by an orbit with
initial conditions in the region of pch than another with initial conditions in the
region of pchb. That is probably the reason why orbit pch could not cross that barrier,
even after $10^{10}$ iterations, while orbit pchc could. Anyway, it was not an easy task
for the latter either, it could do the crossing only after about $3.75 \times 10^{9}$
iterations. 

A caveat is necessary here. We should recall that the trajectories obtained for the same
chaotic orbit with different hardware or software are very different. In fact, although for
regular orbits we obtained essentially the same trajectories with our double and quadruple
precision programs, that was not the case for the partially chaotic orbits. Therefore,
anyone who tries to reproduce our results will find that the orbits we give as regular are
regular, and that those that we give as partially chaotic are partially chaotic. But it is
perfectly possible that, with his computer, he might find that the initial conditions we
give for orbit pchb result in an orbit that invades the region covered by our orbit pch, or
that the initial conditions for orbit pchc give an orbit that only covers the region of our
orbit pchb. However, trying several slightly different initial conditions, he should be able
to find orbits that behave as pchb and pchc.
 
\section{Conclusions}

Using a four dimensional map we have confirmed all the results obtained by \citet{M2017}
with a toy Hamiltonian model with three degrees of freedom. We found partially chaotic
orbits, within a mostly regular domain (see Figure~\ref{wholexy}), that occupy a lane
part of which is shown in Figure~\ref{lanexy}. That lane is made up of curves
corresponding to orbits with different values of an integral of motion. Extrapolating
those curves we found regular orbits that bound the partially chaotic orbit in
question. That is, with a different model we have provided further
evidence that partially chaotic
orbits exist in cocoons well isolated from the Arnold web by regular orbits.
Besides, we found another partially chaotic orbit inside one of the holes
of the first one and a regular orbit that separates them. Finally, inside the first
partially chaotic orbit, we found a broken regular orbit (a cantori) that poses a
porous barrier that hampers the access of that orbit to another partially chaotic
domain, i.e., an example of stickiness.

We have emphasized, both in our previous paper and in the present one, that numerical
results such as ours are valid only over the time spam covered by the numerical
integrations or iterations performed. In that sense, we have here extended the validity
of our previous conclusions to an interval 50 times longer, about 50 million Hubble times
in a galactic context. 

Although the preceeding statement is what strict logic dictates, let us speculate
in this concluding remarks whether our results can be valid for longer time spans.
The big question is: Why not? Here the spectre of Arnold diffusion comes to haunt
us, as we know that it is an extremely slow process. But, in that case, there is
a good reason for that slowness: the 5-D fully chaotic orbits have to find their
way through the insterstices left by the 3-D regular orbits. Here, instead, we
have 4-D partially chaotic orbits surrounded by the 3-D tori of regular orbits. How
would they escape, no matter how long the time at their disposal? The single answer
we can find is that perhaps, somehow, the partially chaotic 4-D orbits transform
into fully chaotic 5-D orbits and then can escape from their 3-D prison. But, to
that, we can pose another question: is there any mechanism that can make that an
orbit that obeys an integral of motion, and for times as long as we have probed,
to cease to obey it? Besides, despite the extremely fine grids investigated and the
very long integration times (or very large numbers of iterations) used to obtain the
LEs of sample orbits, we could find no fully chaotic orbits in the regions of the
lanes of partially chaotic orbits investigated by \citet{M2017} and in the present
work or their surroundings. Therefore, it seems very unlikely that those partially
chaotic orbits might become fully chaotic ones, even for time spans much longer than
those covered by our numerical experiments.

The situation is very different in the frontier between the mostly regular central
domain and the mostly fully chaotic domain that surrounds it in our Figure~\ref{wholexy}
and Figure 1 of \citet{M2017}. The structure of that frontier is quite complex, perhaps
of a fractal nature and very different from the clear simple lanes studied in both
papers, with regular, partially and fully chaotic orbits intermingled. Besides, we have
found that is not unusual for partially chaotic orbits in those regions to reveal a
fully chaotic nature when the integration of their orbits is pursued for longer
times. It is in these regions where one should investigate whether 4-D partially chaotic
orbits place hurdles to 5-D fully chaotic ones but, as already indicated by \citet{M2017},
this will not be an easy task.

Let us finish recalling that partially chaotic orbits are nothing misterious. They are,
in fact, the single chaotic orbits present in Hamiltonian systems with three degrees
of freedom that have an additional integral besides energy, e.g., in systems with
rotational symmetry that conserve the angular momentum component parallel to the
axis of symmetry. Of course, one can argue that those cases can be reduced to systems
with two degrees of symmetry, e.g., studying the motion in the meridional plane in
systems with rotational symmetry. But, then, does it not happen the same with our
orbits? Although we do not know which is the integral that they obey, the 3-D Poincar\'e
maps that we obtained are very similar to the usual Poincar\'e maps for Hamiltonian
systems with two degrees of freedom and, besides, we have found that the phenomenon
of stickiness is also present here. We leave the question open, but we plan to continue
investigating this very interesting problem.

\section*{Acknowledgements}

We are very grateful to D. Pfenniger for the use of his code,
and to H.R. Viturro for his assistance. The comments
of an anonymous reviewer were very useful to improve the original version
of this paper and are gratefully acknowledged. This work
was supported with grants from the Consejo Nacional de Investigaciones
Cient\'{\i}ficas y T\'ecnicas de la Rep\'ublica Argentina, the Agencia
Nacional de Promoci\'on Cient\'{\i}fica y Tecnol\'ogica and the Universidad
Nacional de La Plata.


\bsp	
\label{lastpage}
\end{document}